\documentclass[11pt]{article}
\usepackage[top=1in,bottom=1in,left=1in,right=1in]{geometry}
\usepackage[bf,small,pagestyles]{titlesec}
\usepackage{graphicx}
\usepackage{float}
\usepackage{multirow}
\usepackage{booktabs}
\usepackage{caption}
\usepackage{indentfirst}
\usepackage{subcaption}
\usepackage{amssymb}
\usepackage{amsmath}
\usepackage{cases}
\usepackage{algorithm} 
\usepackage{algpseudocode} 
\usepackage{caption}
\usepackage{hyperref}
\usepackage[figure,table]{hypcap} 
\hypersetup{
	bookmarksnumbered,
	pdfstartview={FitH},
	citecolor={black},
	linkcolor={black},
	urlcolor={black},
	pdfpagemode={UseOutlines}
}
\usepackage{multicol}

\newtheorem{Remark}{Remark}
\usepackage{ulem} 
\usepackage{authblk}

\begin{document}
\title{\textbf{Deep Feynman-Kac Methods for High-dimensional Semilinear Parabolic Equations: Revisit}}

\author[1]{Xiaotao Zheng\thanks{Email: \texttt{20234013002@stu.suda.edu.cn}}}
\author[2]{Xingye Yue\thanks{Corresponding author: \texttt{xyyue@suda.edu.cn}}}
\author[3]{Jiyang Shi}

\affil[1]{Center for Financial Engineering, Soochow University, Suzhou 215008, Jiangsu, China}
\affil[2]{Center for Financial Engineering, Soochow University, Suzhou 215008, Jiangsu, China}

\date{}
\phantomsection
\addcontentsline{toc}{section}{Title}
\maketitle
	
\phantomsection
\addcontentsline{toc}{section}{Abstract}

\begin{abstract}

Deep Feynman-Kac method was first introduced to solve parabolic partial differential equations(PDE) by Beck et al. (SISC, V.43, 2021), named Deep Splitting method since they trained the Neural Networks step by step in the time direction. 
In this paper, we propose a new training approach with two different features. Firstly, neural networks are trained at all time steps globally, instead of step by step. Secondly, the training data are generated in a new way, in which the method is consistent with a direct Monte Carlo scheme when dealing with a linear parabolic PDE. Numerical examples show that our method has significant improvement both in efficiency and accuracy. 

\end{abstract}

\section{Introduction}
With the development of fields such as finance, physics, and medicine, modeling complex problems by high-dimensional partial differential equations (PDEs) has become increasingly important. However, classical numerical methods suffer from the "curse of dimensionality," referring to the exponentially increasing computational complexity associated with increasing dimensions \cite{bellman}.
One potential approach to address this challenge is to leverage the attractive approximation capabilities of neural networks for highly nonlinear functions \cite{weishu1}. As a result, deep learning methods have gained popularity in solving the PDEs. In 2019, Raissi et al. introduced the Physics-Informed Neural Networks (PINNs) method and demonstrated its efficiency for solving the Schrödinger, Burgers, and Allen-Cahn equations \cite{PINN, PINN_2024_1, PINN_2024_2}. 
The method's generalization error was estimated through the training error and number of training samples \cite{Shin, Mishra}.
In addition to PINNs, various deep learning-based approaches have been proposed for solving high-dimensional PDEs, including the Deep Ritz Method \cite{Ritz, Ritz_2024}, the Deep Galerkin Method (DGM) \cite{DGM, DGM_2024}, the Weak Adversarial Network (WAN) \cite{WAN}, the Deep Nitsche Method \cite{nitsche}, and the Deep Finite Volume Method \cite{DFVM}.
However, efficient sampling strategies remain a major challenge, particularly in extremely high-dimensional problems (e.g., $d=100, 500, 1000$).

Certain deep learning techniques based on stochastic processes have been found to be useful for high-dimensional semilinear PDEs.
In 2017, E et al. proposed the Deep Backward Stochastic Differential Equation (DBSDE) method \cite{E2017}. By adapting the PDEs solution to the backward
stochastic differential equations (BSDEs) system \cite{Peng1, Peng2}, 
the core idea of this approach is to reformulate the BSDE as a stochastic optimal control problem and then approximate the control process \( Z \) using deep neural networks.
Furthermore, Han et al. established the theoretical convergence of this method, proving that its error vanishes under the general approximation capabilities of neural networks \cite{Convergence-of-the-deep-BSDE-method}. The effectiveness of the Deep BSDE method has been further demonstrated in various applications, particularly in computational finance, where it has shown strong empirical performance \cite{deepbsde-finance, deepbsde-finance1, deepbsde-finance2}.
In 2021, Beck et al. proposed the Deep Splitting Method (DS) \cite{Beck2} as an alternative approach to solving high-dimensional partial differential equations (PDE).
The method takes advantage of the Feynman-Kac formula \cite{Peng-linear-FK} to represent PDE solutions as conditional expectations. 
The name 'Deep Splitting' comes from the fact that the neural networks are trained step by step in the time direction. 
The DS method has been successfully applied to problems such as the pricing of high-dimensional American options \cite{Beck2, Goudenege, Lapeyre}.
In response to these developments, we propose the Deep Feynman-Kac method under Global Training (DFK-GT) in Section \ref{Sec:DBFK-GT}. Notably, this method recovers the classical Monte Carlo approach when the underlying equation is linear. Furthermore, as demonstrated in Section \ref{Sec:result}, numerical experiments indicate that the DFK-GT method achieves a higher efficiency compared to both the DBSDE and DS method.

The structure of this paper is as follows.
In Section \ref{Sec:theory}, we introduce the backward stochastic differential equations (BSDEs) theory and the Feynman-Kac formula, which are related to the semilinear parabolic PDEs. In Section \ref{Sec:DS_revision}, we revisit the Deep Feynman-Kac method (DFK), and propose a new training approach for it.  In Section \ref{Sec:result}, numerical experiments are presented to compare the performance of different methods for solving high-dimensional semilinear parabolic differential equations. 

\section{Backward Stochastic Differential Equations and Feynman-Kac Formula} \label{Sec:theory}
In this section,
we begin by introducing the Feynman-Kac formula and the connection between the semilinear parabolic partial differential equations (PDE) and the backward stochastic differential equations (BSDEs).

\subsection{Semilinear Parabolic Partial Differential Equations }
We consider the semilinear parabolic PDEs, which can be expressed as: 
\begin{equation}\label{bxxpwxpde}
	\left\{ \begin{aligned}
		& \frac{{\partial u}}{{\partial t}}(t,x)  + \frac{1}{2}{\mathop{\rm Tr}\nolimits} \left( {\sigma {\sigma ^{{T}}}(t,x)\left( {{{{\mathop{\rm Hess}\nolimits} }_x}u} \right)(t,x)} \right) + \nabla u(t,x) \cdot \mu (t,x)  \\
		&\qquad  + f\left( {t,x,u(t,x),{\sigma ^{{T}}}(t,x)\nabla u(t,x)} \right) = 0,\quad (t,x) \in [0,T) \times {R^d},\\
		& u(T,x) = g(x),\quad x \in {R^d}.
	\end{aligned} \right.
\end{equation}
The variables $t$ and $x$ represent respectively time variable and $d$-dimensional spatial variable, $d \gg 1$. $\mu (t,x)$ is a known vector-valued function. $\sigma (t,x)$ is a known $d\times d$ matrix-valued function, and $\sigma^{T}(t,x)$ represents the transpose of $\sigma(t,x)$. $\nabla u(t,x)$ and ${{\mathop{\rm Hess}\nolimits} _x}u(t,x)$ respectively represent the gradient and Hessian of $u(t,x)$ with respect to $x$. $f$ is a known nonlinear function. The objective  is to solve PDEs (\ref{bxxpwxpde}) for $u(0,\xi)$,  $\xi\in{R^d}$.

\subsection{Backward Stochastic Differential Equation (BSDE)}
Here, we follow the approach of Peng \cite{Peng1, Peng2} and references therein to establish the correlation between PDEs(\ref{bxxpwxpde}) and well-posed backward stochastic differential equations (BSDEs).

\textbf{Lemma $2.1$:}
Assuming $(\Omega,\mathcal{F},\mathbb{P})$ be a probability space, make $W=\left(W^{(1)}, \ldots, W^{(d)}\right):[0, T] \times \Omega \rightarrow \mathbb{R}^d$ be a standard Brownian motion, $\mathcal{F}_t$ be an non-decreasing filtration generated by $W$. Let  $X=\left(X^{(1)}, \ldots, X^{(d)}\right):[0, T] \times \Omega \rightarrow \mathbb{R}^d, Y:[0, T] \times \Omega \rightarrow \mathbb{R}, Z:[0, T] \times \Omega \rightarrow \mathbb{R}^d$ be $\mathcal{F}$-adapted stochastic processes. Consider the following BSDE:
\begin{numcases}{}
	X_t=\xi+\int_0^t \mu\left(s, X_s\right) \mathrm{d} s+\int_0^t \sigma\left(s, X_s\right) \mathrm{d} W_s,  \label{Xt}\\
	Y_t=g\left(X_T\right)+\int_t^T f\left(s, X_s, Y_s, Z_s\right) \mathrm{d} s-\int_t^T\left(Z_s\right)^T \mathrm{~d} W_s. \label{Yt}
\end{numcases}
Under certain conditions, the BSDE is well-posed and related to the PDE (\ref{bxxpwxpde}). Specifically, for any $t \in[0, T]$, the following equation holds almost everywhere in probability:
\begin{equation}\label{YZ}
	\left\{ {\begin{aligned}
			& {{Y_t} = u\left( {t,{X_t}} \right),}\\
			& {Z_t} = {\sigma ^ {T} }\left( {t,{X_t}} \right)\nabla u\left(t,X_t \right).
	\end{aligned}} \right.
\end{equation}
In other words, if the $d$-dimensional stochastic process $\left\{X_t\right\}_{t \in[0, T]}$ satisfies equation(\ref{Xt}),
then the solution of PDE (\ref{bxxpwxpde}) satisfies the following stochastic differential equation (SDE):

\begin{equation}\label{BSDEeq}
	\begin{aligned}
		u(t,{X_t}) - u(0,\xi) & =  - \int_0^t {f\left( {s,X_s,u(s,X_s),{\sigma ^{{T}}}(s,X_s)\nabla u(s,X_s)} \right)} ds \\
		& +\int_0^t {{{[\nabla u(s,X_s)]}^T}\sigma (s,X_s)d{W_s}} .
	\end{aligned}
\end{equation}

\subsection{Feynman-Kac Formula}
\textbf{Lemma $2.2$:} 
If the solution $u(t, x)$ to the PDE (\ref{bxxpwxpde}) belongs to $C^{1,2}\left([0,T]\times \mathbb{R}^d\right)$, then it can be represented as a conditional expectation,
\begin{equation}\label{F-Keq}
	u(t,x) = \mathbb{E}\left[ {g\left( {{X_T}} \right) + \int_t^T f \left( {s,{X_s},u\left( {s,{X_s}} \right),{\sigma ^{{T}}}(s,{X_s})\nabla u\left( {s,{X_s}} \right)} \right){\rm{d}}s{\rm{\mid }}{X_t} = x} \right].
\end{equation}
The $d$ dimensional stochastic process $\left\{X_s\right\}_{s \in[t, T]}$ satisfies the SDE:
\begin{equation}\label{2.7}
		\left\{ {\begin{aligned}
				&{d{X_s} = \mu \left( {s,{X_s}} \right)ds + \sigma \left( {s,{X_s}} \right){\rm{d}}{W_s}}\\
				&{{X_t} = x }
		\end{aligned}} \right.,
\end{equation}
where $W_s \in \mathbb{R}^d$ is a standard Brownian motion.

\section{Deep Feynman-Kac Method} \label{Sec:DS_revision}
In this section, we will first provide an overview of the Deep Splitting (DS) method, as presented in the work of Beck \cite{Beck2}. Then we will consider the Deep Splitting method under global training and the Deep Feynman-Kac method under Global Training (DFK-GT).

\subsection{Deep Splitting (DS) Method}
The DS method is an efficient deep learning technique for solving PDEs (\ref{bxxpwxpde}) based on the Feynman-Kac formula. This method employs a splitting scheme to decompose the original PDE into multiple subproblems, which can be solved more easily.

We begin by discretizing the time domain \([0,T]\) as  
\[
0 = t_0 < t_1 < \cdots < t_N = T,
\]  
where the time step is given by \(\Delta t_n = t_n - t_{n-1}\). Using this discretization, the Feynman-Kac formula \eqref{F-Keq} is applied iteratively in a backward time-stepping manner, leading to the following recursion: 
\begin{equation}\label{F-Kmodel}
	\left\{ \begin{aligned}
		&u(T,x) = g(x),\\
		&u\left( {t_{n-1},x} \right) = \mathbb{E}\left[ u\left( {t_{n}, X_{t_{n}}} \right) + \int_{t_{n-1}}^{t_{n}} f\left(s,X_s,u\left(s,X_s\right),z\left(s,X_s\right)\right) ds \mid X_{t_{n-1}}=x \right],\\
		&z(t,x) = {\sigma^T}\left(t,x\right)\nabla u\left(t,x\right),
	\end{aligned} \right.
\end{equation}
where $\left\{X_s\right\}_{s \in[t_{n-1}, t_{n}]}$ satisfies the SDE \eqref{2.7} and ${n = 1,2, \ldots ,N }$. 
To facilitate numerical implementation, we approximate the integral term in equation \eqref{F-Kmodel} using a quadrature rule,
\begin{equation}\label{shuzhijifen}
	\begin{aligned}
		&\int_{{t_{n-1}}}^{{t_{n}}} {f\left( {s,{X_s},u\left( {s,{X_s}} \right),{\sigma ^{{T}}}\left( {s,{X_s}} \right)\nabla u\left( {s,{X_s}} \right)} \right)} {\rm{d}}s\\
		\approx &f\left( {{t_{n }},{X_{{t_n}}},u\left( {{t_{n }},{X_{{t_{n }}}}} \right),{\sigma ^{{T}}}\left( {{t_{n }},{X_{{t_{n }}}}} \right)\nabla u\left( {{t_{n }},{X_{{t_{n }}}}} \right)} \right) \cdot \Delta {t_n}.
	\end{aligned}
\end{equation}
Meanwhile, \(M\) independent sample paths \(\{X_{t_n}^m\}_{0 \leq n \leq N, m = 1, \dots, M}\) are generated by simulating the SDE \eqref{2.7},
\begin{equation}
    \begin{aligned}
        X_{t_n}^m = X_{t_{n-1}}^m + \mu\left(t_{n-1}, X_{t_{n-1}}^m \right) \Delta t_n + \sigma\left(t_{n-1}, X_{t_{n-1}}^m \right) \Delta W_n, \quad X_0 = \xi.
    \end{aligned}
\end{equation}  
Thus, model (\ref{F-Kmodel}) can be expressed as follows:
\begin{equation}\label{4.5}
		\left\{ \begin{aligned}
			&X_{t_n}^m = X_{t_{n-1}}^m + \mu\left(t_{n-1}, X_{t_{n-1}}^m \right) \Delta t_n + \sigma\left(t_{n-1}, X_{t_{n-1}}^m \right) \Delta W_n, \quad X_0 = \xi,\\
			&u\left( {{t_{n-1}},x} \right) = \mathbb{E}\left[
			u\left( {{t_{n }},{X_{{t_{n }}}^m}} \right)  + f\left( {{t_{n}},{X_{{t_{n }}}^m},u\left( {{t_{n }},{X_{{t_{n }}}^m}} \right),z\left( {{t_{n }},{X_{{t_{n }}}^m}} \right)} \right) \Delta {t_n}{\rm{\mid }}{X_{{t_{n-1}}}^m} = x
			\right]\!,\\
			&z(t,x) = {\sigma ^T}\left( {t,x} \right)\nabla u\left( {t,x} \right), \quad u(T,x) = g(x),
		\end{aligned} \right.
\end{equation}
where ${n = 1,2, \ldots ,N, m = 1, \dots, M }$.

Instead of using least-squares regression \cite{LSMC}, the conditional expectation $u\left( {{t_{n-1}},{X_{{t_{n-1}}}}} \right)$ 
is approximated using a neural network, denoted as ${u^{{\theta _{n - 1}}}}\left( {{X_{{t_{n - 1}}}}} \right)$: ${\mathbb{R}^d}\to\mathbb{R}$. 
To approximate the gradient \(\nabla u(t_{n-1}, X_{t_{n-1}})\), we leverage the automatic differentiation capabilities of TensorFlow \cite{tensorflow} to compute \(\nabla u^{\theta_{n-1}}(X_{t_{n-1}})\): \(\mathbb{R}^d \to \mathbb{R}^d\). 
Next, the neural network \(u^{\theta_{n-1}}(X_{t_{n-1}})\) is then trained at each time step using the data pairs $\left\{ {X_{{t_{n-1}}}^m,\hat u\left( {{t_{n-1}},X_{{t_{n-1}}}^m} \right)} \right\}$.
Here, ${\hat u\left( {{t_{n-1}},X_{{t_{n-1}}}^m} \right)}$ is a realization of the value function ${ u\left( {{t_{n-1}},X_{{t_{n-1}}}^m} \right)}$.In detail,
\begin{equation}\label{TS-FK-LSMC-BECK}
	\left\{ {\begin{aligned}
            & X_{t_n}^m = X_{t_{n-1}}^m + \mu\left(t_{n-1}, X_{t_{n-1}}^m \right) \Delta t_n + \sigma\left(t_{n-1}, X_{t_{n-1}}^m \right) \Delta W_n, \quad X_0 = \xi, \\
		  &{\hat u({t_{n-1}},X_{{t_{n-1}}}^m) = {u^\theta }({t_{n }},X_{{t_{n}}}^m) + f({t_{n}},X_{{t_{n}}}^m,{u^\theta }({t_{n }},X_{{t_{n}}}^m),{z^\theta }({t_{n}},X_{{t_{n}}}^m)) \Delta {t_n},}\\
		  &{z^\theta }(t,x) = {\sigma ^T}(t,x)\nabla {u^\theta }(t,x),\\
            &{\hat u({t_{N - 1}},X_{{t_{N - 1}}}^m) = g(X_T^m) + f(T,X_T^m,g(X_T^m),{\sigma ^T}(T,X_T^m)\nabla g(X_T^m)) \Delta {t_N},}
	\end{aligned}} \right.
\end{equation}
Therefore, we define the loss function at time step ${{t_{n-1}}}$:
\begin{equation}\label{TS-FK-LSMC-LOSS-BECK}
	{l_{n - 1}}({\theta _{n - 1}}) = \frac{1}{M}\sum\limits_{m = 1}^M {{{\left( {{u^{{\theta _{n - 1}}}}(X_{{t_{n - 1}}}^m) - \hat u({t_{n - 1}},X_{{t_{n - 1}}}^m)} \right)}^2}}  ,
\end{equation}
where $n$ is in descending order from $N,N-1,...,2$. After training the network parameters ${\theta _{n-1}}$ step by step using Adam optimizer, the approximate solution $\hat u(0,\xi )$ can be obtained via equation (\ref{4.7.2.1}). 
\begin{equation}\label{4.7.2.1}
	\hat u(0,\xi ) = \mathbb{E}\left[ {{u^\theta}\left( {{t_1},X_{{t_1}}^m} \right) \! + \!  f\left( {{t_1},X_{{t_1}}^m,{u^\theta}\left( {{t_1},X_{{t_1}}^m} \right),{\sigma ^T}\left( {{t_1},X_{{t_1}}^m} \right)\nabla {u^\theta}\left( {{t_1},X_{{t_1}}^m} \right)} \right) \cdot \Delta {t_0}{\rm{\mid }}{X_0} = \xi } \right].
\end{equation}

\subsection{Deep Splitting method under Global Trainning (DS-GT)}
In this subsection, we extend the Deep Splitting (DS) method by introducing global training. 
Unlike the conventional stepwise training approach, which optimizes neural network parameters at each time step independently, global training considers the entire time domain simultaneously during the training process.
This means that during the optimization of neural network parameters, interactions between different time intervals are taken into account. By leveraging global training, our aim is to enhance both accuracy and computational efficiency.

The Deep Splitting method under Global Training also solves the model (\ref{4.5}), where $\theta_{n-1}$ represents the network parameters used for approximating the conditional expectation
$u\left( {{t_{n-1}},{X_{{t_{n-1}}}}} \right)$ with the neural network ${u^{{\theta _{n - 1}}}}\left( {{X_{{t_{n - 1}}}}} \right)$. 
To facilitate global training, \(N-1\) fully connected feedforward neural networks are linked, and the collective set of network parameters is denoted as  
\(\theta=\left\{\theta_1, \theta_2, \ldots, \theta_{N-1}\right\}\).  
The loss function is then redefined as the cumulative sum of individual loss functions across all time steps:  
\begin{equation}\label{DS-GT-LOSS}
	l(\theta ) = \sum\limits_{n = 2}^N {{l_{n - 1}}({\theta _{n - 1}})},
\end{equation}
where ${l_{n - 1}}({\theta _{n - 1}})$ is defined in Equation(\ref{TS-FK-LSMC-LOSS-BECK}).
Having optimized the network parameters $\theta$ with the Adam algorithm, we can obtain the approximate solution $\hat u(0,\xi )$ through equation (\ref{4.7.2.1}). 

\subsection{Deep Feynman-Kac method under Global Training (DFK-GT)} \label{Sec:DBFK-GT}
Building upon the DS-GT method, we propose the Deep Feynman-Kac method under Global Training (DFK-GT). This approach aims to further improve the accuracy of the numerical approximation by refining the training data pairs used in the network training process. It is worth noting that if the PDE (\ref{bxxpwxpde}) reduces to a linear equation, this method is equal to the general Monte Carlo method.

After obtaining the model (\ref{4.5}), we employ the same approach to approximate the conditional expectation $u\left( {{t_{n-1}},{X_{{t_{n-1}}}}} \right)$ 
by a neural network ${u^{{\theta _{n - 1}}}}\left( {{X_{{t_{n - 1}}}}} \right)$, approximate the gradient function $\nabla u\left( {{t_{n - 1}},{X_{{t_{n - 1}}}}} \right)$ with $\nabla {u^{{\theta _{n - 1}}}}\left( {{X_{{t_{n - 1}}}}} \right)$: ${\mathbb{R}^d} \to {\mathbb{R}^d}$. 
However, unlike the Deep Splitting method, we train  ${u^{{\theta _{n - 1}}}}\left( {{X_{{t_{n - 1}}}}} \right)$ at every time step using a set of modified data pairs $\left\{ {X_{{t_{n-1}}}^m,\tilde u\left( {{t_{n-1}},X_{{t_{n-1}}}^m} \right)} \right\}$. Here, ${\tilde u\left( {{t_{n-1}},X_{{t_{n-1}}}^m} \right)}$ is an another realisation of the value function ${ u\left( {{t_{n-1}},X_{{t_{n-1}}}^m} \right)}$, thereby deriving a new solution scheme:
\begin{equation}\label{TS-FK-LSMC-LOSS1}
	\left\{ {\begin{aligned}
                & X_{t_n}^m = X_{t_{n-1}}^m + \mu\left(t_{n-1}, X_{t_{n-1}}^m \right) \Delta t_n + \sigma\left(t_{n-1}, X_{t_{n-1}}^m \right) \Delta W_n, \quad X_0 = \xi, \\    
			&{\tilde u({t_{n-1}},X_{{t_{n-1}}}^m) = \tilde u({t_{n}},X_{{t_{n }}}^m) + f({t_{n }},X_{{t_{n }}}^m,{u^\theta }({t_{n }},X_{{t_{n }}}^m),{z^\theta }({t_{n }},X_{{t_{n }}}^m)) \Delta {t_n},}\\
			&{z^\theta }(t,x) = {\sigma ^T}(t,x)\nabla {u^\theta }(t,x),\\
                &{\tilde u({t_{N - 1}},X_{{t_{N - 1}}}^m) = g(X_T^m) + f(T,X_T^m,g(X_T^m),{\sigma ^T}(T,X_T^m)\nabla g(X_T^m)) \Delta {t_n},}
	\end{aligned}} \right.
\end{equation}
The loss function is redefined as
\begin{equation}\label{OT-FK-LSMC-LOSS}
	l(\theta ) = \sum\limits_{n = 2}^N {{l_{n - 1}}({\theta _{n - 1}})} ,
\end{equation}
with
\begin{equation}\label{TS-FK-LSMC-LOSS}
	{l_{n - 1}}({\theta _{n - 1}}) = \frac{1}{M}\sum\limits_{m = 1}^M {{{\left( {{u^{{\theta _{n - 1}}}}(X_{{t_{n - 1}}}^m) - \tilde u({t_{n - 1}},X_{{t_{n - 1}}}^m)} \right)}^2}}  .
\end{equation}
After training the network parameters $\theta=\left\{\theta_1, \theta_2, \ldots, \theta_{N-1}\right\}$ using Adam optimizer, we obtain the approximate solution $\tilde u(0,\xi )$ via equation (\ref{4.7.2}). 
\begin{equation}\label{4.7.2}
	\tilde u(0, \xi) = \mathbb{E}\left[ {\tilde u\left( {{t_1},X_{{t_1}}^m} \right) + f\left( {{t_1},X_{{t_1}}^m,{u^\theta}\left( {{t_1},X_{{t_1}}^m} \right),{\sigma ^T}\left( {{t_1},X_{{t_1}}^m} \right)\nabla {u^\theta}\left( {{t_1},X_{{t_1}}^m} \right)} \right) \cdot \Delta {t_0}{\rm{\mid }}{X_0} = \xi } \right].
\end{equation}

\begin{figure*}[!hbtp]
	\begin{center}
		\includegraphics[angle=0,width=6in]{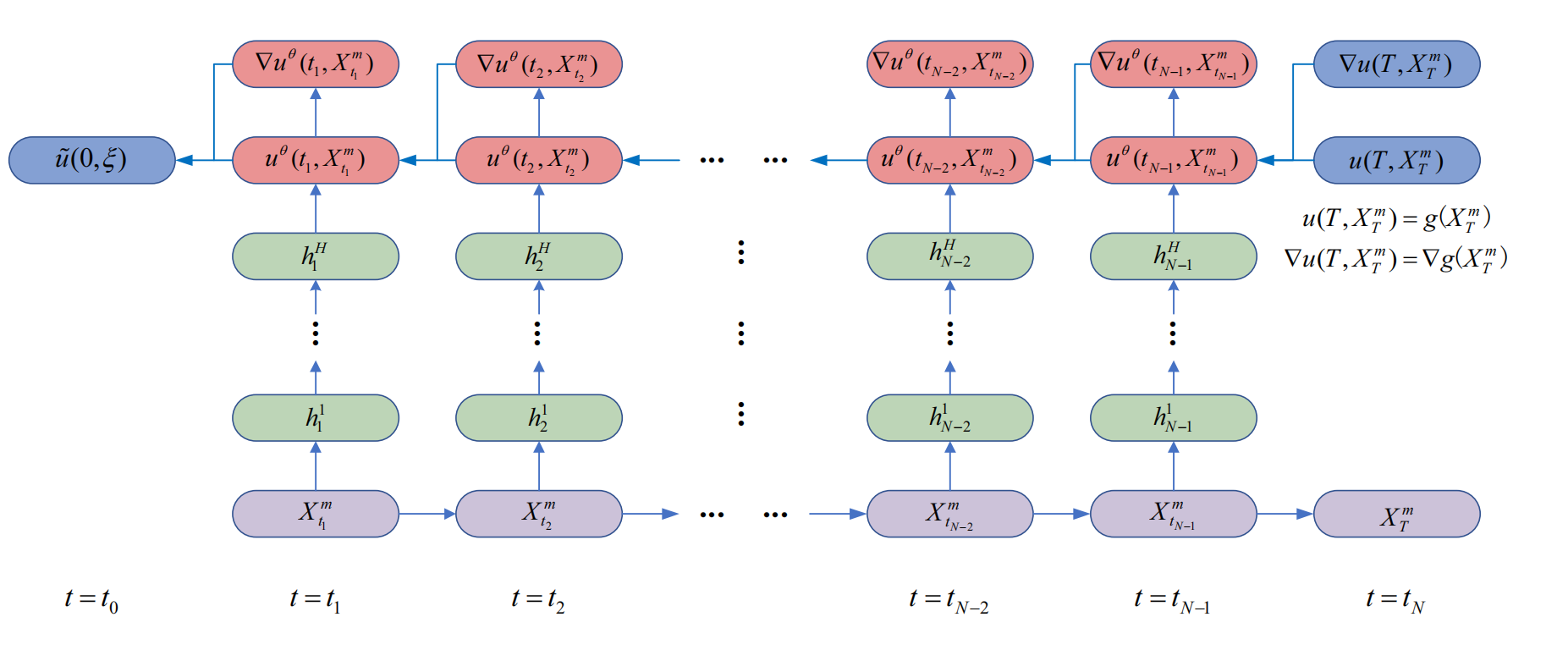}
	\end{center}
	\caption{Deep Feynman-Kac Algorithm under Global Training. Here, we also fully connect $N-1$ sub-networks and train them as a whole. The structure of the sub-networks is the same as that in the Deep BSDE method. However, unlike that, the network here directly approximates the value function $u^\theta (t_n,X_{t_n}^m)$ instead of its gradient $\nabla u^\theta (t_n,X_{t_n}^m)$. As a result, we reduce two network parameters, namely ${\theta _{{u_0}}}$ and ${\theta _{\nabla {u_0}}}$.}
	\label{fig1}
\end{figure*}

It can be observed that the method can be simplified to a general Monte Carlo algorithm, if the PDE(\ref{bxxpwxpde}) reduces to a linear form.
Having optimized the network parameters $\theta$ with the Adam algorithm, we can obtain the approximate solution $\tilde u(0, \xi)$ through equation (\ref{4.7.2}). 
For more details about the method, readers can refer to Algorithm \ref{alg:deep_backward_Feynman-Kac2_1} and Figure \ref{fig1}.

\begin{algorithm}
	\caption{Deep Feynman-Kac method under Global Training}
	\label{alg:deep_backward_Feynman-Kac2_1}
	\begin{algorithmic}[1]
		\Require $N, M, T, \Delta t_n, It, B$
		\State Time partition: $0 = t_0 < t_1 < \cdots < t_N = T$, $\Delta t_n = t_{n} - t_{n - 1}$
		\State Sample Brownian motion increment $\Delta W_{t_n}^{(m)}$, get paths $X_{t_n}^{(m)}$, $m = 1,2,\dots,M$, $n = 0,1,\dots,N$
		\State Initialize neural network parameters $\theta$
		\For {$i \gets 1$ to $It$}
		\State Sample $B$ paths $X_{t_n}^{(m)}$, $m = 1,2,\dots,B$, $n = 0,1,\dots,N$
		\State Calculate $u(T,X_T^{(m)}) = g(X_T^{(m)})$, $m = 1,2,\dots,B$
		\For {$n \gets N$ to $2$}
		\State Approximating the value function  using neural networks:
		\State ${u^\theta }({t_{n}},X_{{t_{n}}}^m)  \approx  u({t_{n}},X_{{t_{n}}}^m)$,$\nabla {u^\theta }({t_{n}},X_{{t_{n}}}^m) \approx \nabla u({t_{n}},X_{{t_{n}}}^m)$, $m = 1,2,\dots,B$
		\State Approximating the conditional expectation using neural networks to calculate the value 
		\State function at the $n-1^{th}$ time step:
		\State ${u}\left( {{t_{n - 1}},x} \right) = \mathbb{E}\left[ {\begin{array}{*{20}{l}}
				{u\left( {{t_{n}},X_{{t_{n}}}^m} \right) + F_{n}\left| {X_{{t_{n - 1}}}^m = x} \right.}
		\end{array}} \right],$ where
		\State $F_{n} = f\left( {{t_{n}},X_{{t_{n}}}^m,{u^\theta }\left( {{t_{n}},X_{{t_{n}}}^m} \right),{\sigma ^T}({t_{n}},X_{{t_{n}}}^m)\nabla {u^\theta }({t_{n}},X_{{t_{n}}}^m)} \right) \cdot \Delta {t_n}$, $m = 1,2,\dots,B$
		\EndFor
		\State Calculate loss function $l(\theta) = \frac{1}{B}\sum_{n=2}^{N}\sum_{m=1}^B\left[u^\theta(t_{n-1}, X_{t_{n-1}}^m) - \tilde u(t_{n-1}, X_{t_{n-1}}^m)\right]^2$,where
		\begin{equation*}
			\left\{
			\begin{aligned}
				&{\tilde u({t_{N - 1}},X_{{t_{N - 1}}}^m) = g(X_T^m) + f(T,X_T^m,g(X_T^m),{\sigma ^T}(T,X_T^m)\nabla g(X_T^m)) \cdot \Delta {t_n},}\\
				&{\tilde u({t_{n-1}},X_{{t_{n-1}}}^m) = \tilde u({t_{n}},X_{{t_{n }}}^m) + f({t_{n }},X_{{t_{n }}}^m,{u^\theta }({t_{n }},X_{{t_{n }}}^m),{z^\theta }({t_{n }},X_{{t_{n }}}^m)) \cdot \Delta {t_n},}\\
				&{{z^\theta }(t,x) = {\sigma ^T}(t,x)\nabla {u^\theta }(t,x),n = 2,...,N - 1.}
			\end{aligned}
			\right.
		\end{equation*}
		\State optimize the network parameters $\theta$ with the Adam algorithm 
		\State Get the approximate solution of the equation:
		\EndFor 
		\State $\tilde u(0, \xi) = \mathbb{E}\left[\tilde u(t_1, X_{t_1}^m) + f(t_1, X_{t_1}^m, u^\theta(t_1, X_{t_1}^m), \sigma^T(t_1, X_{t_1}^m)\nabla u^\theta(t_1, X_{t_1}^m))\cdot \Delta t_0\mid X_0 = \xi\right]$ , where
		$m=1,2,...,M$
		\Ensure Approximation of solution based on Feynman-Kac equation
	\end{algorithmic}
\end{algorithm}

\section{Numerical Results} \label{Sec:result}
As discussed in Section 3, the Deep Splitting (DS) method is utilized to decompose the partial differential equation (PDE) (\ref{bxxpwxpde}) into multiple subproblems, which are then recursively solved using neural networks. To further enhance the training process, we introduce the Deep Splitting Method under Global Training (DS-GT) by replacing the step-by-step training approach with global training. This approach shares similarities with the Deep BSDE (DBSDE) method \ref{Appendix:DBSDE}.
Building upon the DS-GT method, we propose the Deep Feynman-Kac method under Global Training (DFK-GT) by modifying the training data pairs used in the network training process. It is interesting to note that when the PDE (\ref{bxxpwxpde}) reduces to a linear equation, this method is equivalent to the general Monte Carlo method. 

In this section, we will proceed to compare the performance of the DBSDE, DS, DS-GT and DFK-GT method in solving the Hamilton-Jacobi-Bellman (HJB) equation, the Allen-Cahn equation, and the PricingDiffRate equation. By evaluating these methods in terms of computational efficiency, accuracy, and robustness, we aim to confirm the significance of global training and training data pair modeling. Ultimately, this comparative analysis will lead to the determination of the optimal solution for solving high-dimensional semilinear parabolic equations (\ref{bxxpwxpde}).

In the DBSDE, DS-GT and DFK-GT methods, we fully connect $N-1$ steps of feedforward neural networks and train them as a whole. In contrast, the DS method follows a step-by-step training approach, as mentioned in Section \ref{Sec:DS_revision}.
In each time step, neural network consists of 4 parts, including one input layer (d-dimensional), two hidden layers (both d+10 dimensional), and one output layer (d-dimensional for the DBSDE method, 1-dimensional for the DS, DS-GT and DFK-GT method). The number of neurons in each hidden layer is d+10. We also adopt batch normalization (BN) \cite{BN} between the affine transformations and activation functions in the fully connected layers, which ensures that the batch data are uniformly distributed before being input to the next layer. This makes it easier for deep neural networks to converge and reduces the risk of overfitting. 
For network optimization, we employ the Adam algorithm \cite{ADAM}, with a total of 512 test samples and a batch size of 256 samples used for each iteration. 
The numerical experiments presented below were performed in Python using the Tensorflow processor on a Dell computer equipped with a 2.30 Gigahertz (GHz) Intel Core i7-11800H.

\subsection{Hamilton-Jacobi-Bellman (HJB) equation}
In this subsection, we apply the DBSDE, DS, DS-GT, DFK-GT methods to numerically solve the Hamilton-Jacobi-Bellman (HJB) equation in 100 dimensions.

Consider a semilinear parabolic PDE with a gradient squared term, following the general form of a semilinear parabolic PDE (\ref{bxxpwxpde}). Let $\mu(t,x) = 0$, $\sigma=\sqrt{2}$, $f(t,x,y,z)=-|z_{\mathbb{R}^d}|^2$, and $g(x) = \ln\left(\frac{1}{2}\left[1+|x|_{\mathbb{R}^d}^2\right]\right)$. Now, We get an equation that satisfies the terminal condition $u(T,x)=g(x)$ and holds for all $t\in[0,T)$ and $x\in{\mathbb{R}^d}$:
\begin{equation}\label{HJBeq}
	\left\{\begin{array}{l}
		\frac{\partial u}{\partial t}(t, x)+\left(\Delta_x u\right)(t, x)-\left\|\left(\nabla_x u\right)(t, x)\right\|_{}^2=0,(t, x) \in[0, T) \times \mathbb{R}^d \\
		u(T, x)=g(x), x \in \mathbb{R}^d
	\end{array}\right.
\end{equation}
Fixed $d=100$, $T=1$, $\xi=(0,0,\ldots, 0)\in\mathbb{R}^d$, the explicit solution to the equation can be obtained through the Cole-Hopf transformation \cite{Cole-Hopf1, Cole-Hopf2}: $u(t, x)=-\ln \left(\mathbb{E}\left[\exp \left(-g\left(x+\sqrt{2} W_{T-t}\right)\right)\right]\right)$, where $u(0,\xi)\approx 4.5901$. 
Based on the loss function (\ref{TS-FK-LSMC-LOSS-BECK}), (\ref{DS-GT-LOSS}), (\ref{OT-FK-LSMC-LOSS}), and (\ref{BSDE-LOSS}), we perform independent calculations five times with the hyper-parameter $\mathrm{N}=20$. 

\begin{figure*}
	\vspace{-1cm}
	\centering
	\begin{minipage}[t]{0.5\linewidth}
		\centering
		\includegraphics[width=\textwidth]{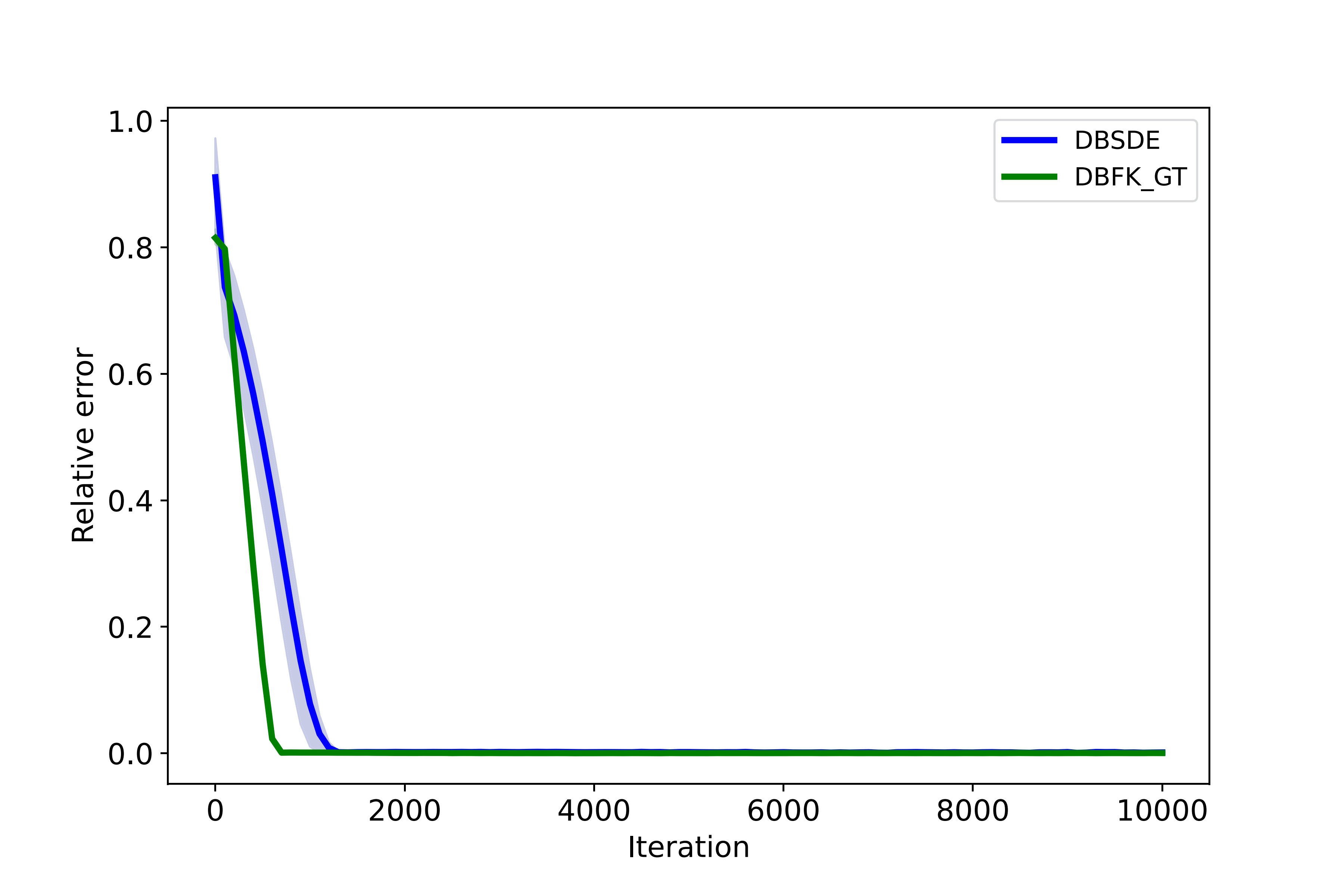}
		\label{fig2:first}
	\end{minipage}%
	\hfill
	\begin{minipage}[t]{0.5\linewidth}
		\centering
		\includegraphics[width=\textwidth]{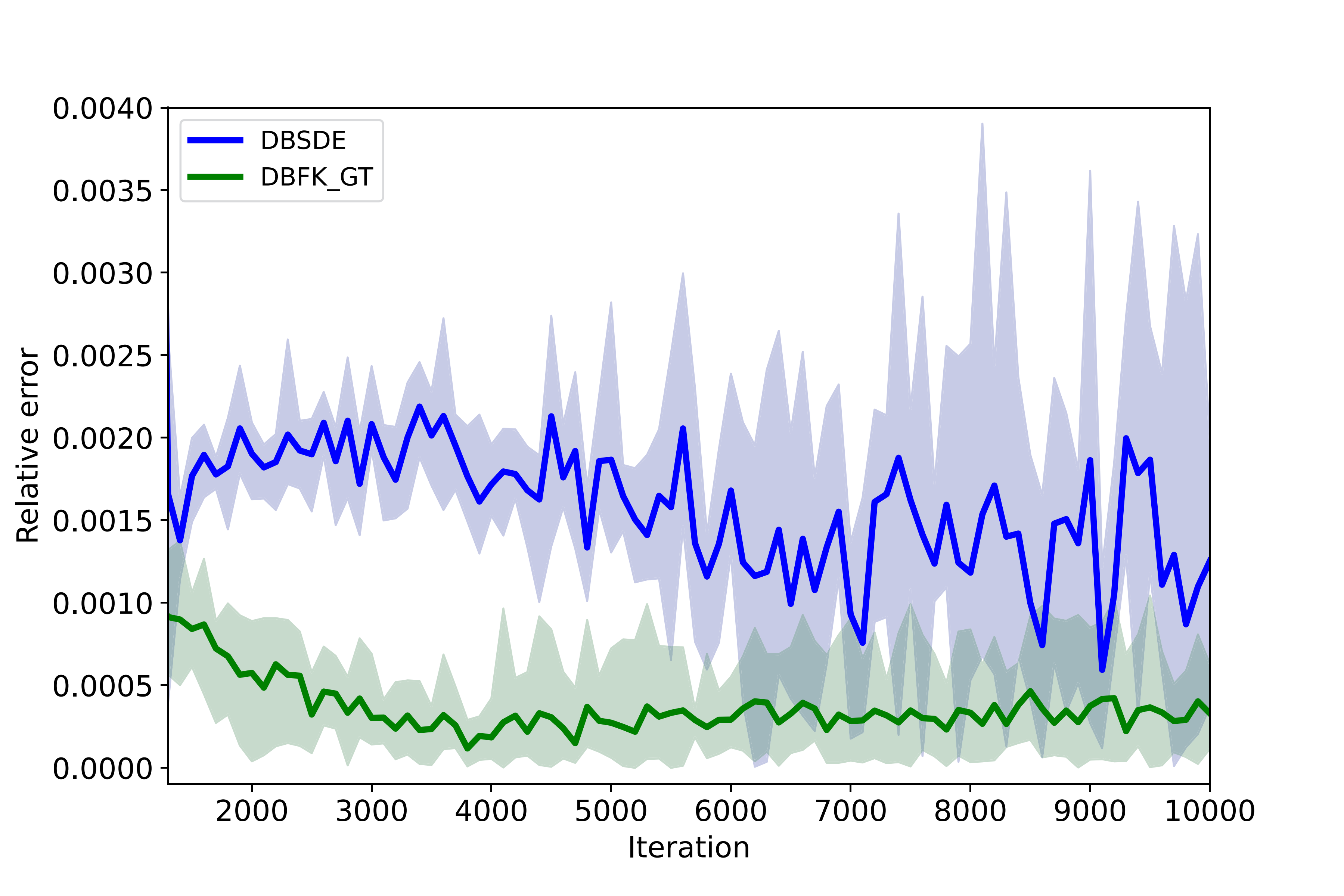}
		\label{fig2:second}
	\end{minipage}
	\centering
	\vspace{-0.7cm}
	\caption{Comparison of Two Methods for Solving the HJB Equation}
	\label{figHJB}
\end{figure*}

The results are reported in Table \ref{biaohjb}. 
It was observed that after sufficient training, the global training scheme yielded higher accuracy compared to the step-by-step training scheme. Further comparing the DS-GT and DFK-GT methods, it is found that re-modeling the data pairs significantly enhanced the efficiency and accuracy of subsequent computations. 
In summary, compared to the DS method, the DFK-GT method exhibits significant improvements in accuracy and robustness.
Moreover, as illustrated in Figure \ref{figHJB}, the DFK-GT method not only surpasses the widely used DBSDE method in terms of accuracy, but also achieves substantial computational efficiency gains.

\begin{table}[htbp]
	\centering
	\caption{Numerical Results for HJB Equation}
	\resizebox{0.8\textwidth}{!}{
		\begin{tabular}{ccccccccc}
			\hline \hline
			&  & \multicolumn{7}{c}{Number of iteration steps} \\
			\cmidrule{3-9}
			&   & 100 & 200 & 600 & 1000 & 2000 & 5000 & 10000 \\
			\midrule
			\multirow{4}{*}{Runtime} & DBSDE  & 11.4 & 15.2 & 32.2 & 49.8 & 91.4 & 216.2 & 423.2 \\
			& DS  & 18.8 & 21.8  & 34.4  & 47.8  & 79.8  & 180.6 & 355.6 \\
			& DS-GT  & 10.2  & 12.2  & 22  & 31.6 & 54.8 & 125.4 & 242.8 \\
			& DFK-GT  & 8.4 & 11.6 & 19.8 & 28 & 51.4 & 120.8 & 229.4 \\
			\midrule
			\multirow{4}{*}{Relative error} & DBSDE  & 7.37e-01 & 6.93e-01 & 4.93e-01 & 7.77e-02 & 1.90e-03 & 1.87e-03 & 1.25e-03 \\
			& DS  & 1.57e-01 & 4.98e-03 & 8.50e-03 & 7.02e-03 & 7.13e-03 & 6.39e-03 & 8.02e-03 \\
			& DS-GT  & 1.03 & 1.02 & 6.34e-01 & 2.08e-01 & 1.03e-03 & 1.48e-03 & 1.86e-03  \\
			& DFK-GT  & 7.97e-01 & 6.28e-01 & 2.30e-02 & 1.07e-03 & 5.75e-04 & \textbf{2.73e-04} & 3.29e-04 \\
			\midrule
			\multirow{4}{*}{Loss function} & DBSDE  & 3.45 & 3.23 & 2.00 & 4.03e-01 & 2.20e-02 & 1.99e-02 & 1.8e-02 \\ 
			& DS  & 2.09e-03 & 1.56e-05 & 4.26e-05 & 2.26e-05 &2.48e-05 & 4.66e-05 & 6.90e-05  \\
			& DS-GT  & 2.90e+01 & 9.53 & 1.16 & 1.09e-01 &2.14e-02 & 4.55e-02 & 4.47e-02 \\
			& DFK-GT  & 8.82e+01 & 6.77e+01 & 4.85 & 4.04e-01 & 3.94e-01 & 4.40e-01 & 4.24e-01  \\
			\hline \hline
		\end{tabular}
	}
	\label{biaohjb}
\end{table}

Importantly, a consistent relationship is observed between variations in the loss function and the accuracy of the methods. Specifically, when the loss function exhibits limited fluctuations (i.e., the loss function ceases to decrease), the optimal number of iterations required for algorithm convergence can be determined. The corresponding results are summarized in Table \ref{biaohjb1}. Notably, all methods yield highly accurate results, with DFK-GT achieving the highest computational efficiency and the lowest standard deviation.

\begin{table}
	\caption{Iteration Count at Loss Convergence for the HJB Equation}
	\centering
	\resizebox{0.6\textwidth}{!}{
		\begin{tabular}{ccccc}
			\hline \hline
			Methods  & Iterations & Relative error & Runtime & Standard deviation \\
			\hline
			DBSDE  & 1800 &1.83e-03 & 83.2 & 8.65e-03 \\
			\hline
			DS & 600 & 8.50e-03 & 34.4 & 1.46e-02 \\
			\hline
			DS-GT  & 1700 &  4.23e-03 & 47.6 & 8.36e-03 \\
			\hline
			DFK-GT   & 1200 & \textbf{1.16e-03} & \textbf{33.0} & \textbf{3.11e-03} \\
			\hline \hline
		\end{tabular}
	}
		\label{biaohjb1}
\end{table}

Additionally, by recording the experimental results where the relative error first falls below \(10^{-2}\), we can ascertain the ideal number of iterations required for algorithm convergence. The corresponding results are presented in Table \ref{biaohjb2}.

\begin{table}
	\caption{Iteration Count for the HJB Equation (Relative Error Below \(10^{-2}\))}
	\centering
	\resizebox{0.6\textwidth}{!}{
		\begin{tabular}{ccccc}
			\hline \hline
			Methods  & Iterations & Relative error & Runtime & Standard deviation \\
			\hline
			DBSDE  & 1200 &8.43e-03 & 83.2 & 1.56e-02 \\
			\hline
			DS & 200 & 4.98e-03 & \textbf{21.8} & 1.80e-02 \\
			\hline
			DS-GT  & 1600 &  9.28e-03 & 45.4 & 1.37e-02 \\
			\hline
			DFK-GT   & 700 & 1.06e-03 & \textbf{26.0} & \textbf{1.03e-03} \\
			\hline \hline
		\end{tabular}
		}
	\label{biaohjb2}
\end{table}
	
\subsection{Allen-Cahn equation}
In this subsection, we test the effectiveness of several methods in solving a 100-dimensional Allen-Cahn PDE with cubic nonlinearity. Referring to the general form of the semilinear parabolic equation (\ref{bxxpwxpde}), we consider the case where $\alpha=1$, $f(y,z) = y - y^3$, and $g(x) = \left[2 + \frac{2}{5} \|x\|_{\mathbb{R}^d}^2\right]^{-1}$. The equation that satisfies the terminal condition $u(T,x) = g(x)$ and holds for all $t \in [0,T)$ and $x \in \mathbb{R}^d$:
\begin{equation}\label{AllenCahneq}
	\left\{\begin{array}{l}
		\frac{\partial u}{\partial t}(t, x)+\left(\Delta_x u\right)(t, x)+u(t, x)-[u(t, x)]^3=0,(t, x) \in[0, T) \times \mathbb{R}^d \\
			u(T, x)=g(x), x \in \mathbb{R}^d
	\end{array}\right.
\end{equation}
	
\begin{figure*}
	\centering
	\begin{minipage}[t]{0.5\linewidth}
		\centering
		\includegraphics[width=\textwidth]{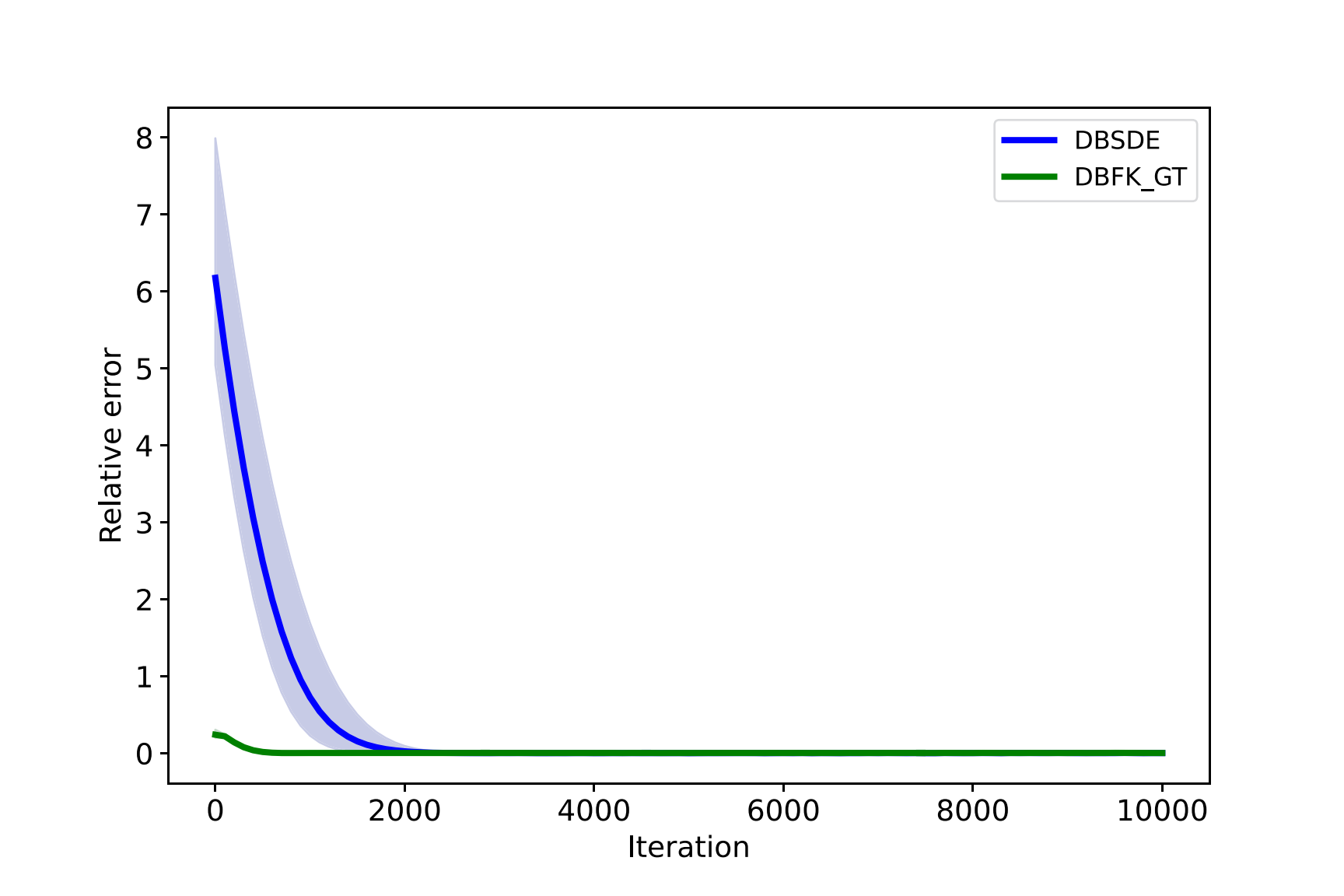}
		\label{fig3:first}
	\end{minipage}%
	\hfill
	\begin{minipage}[t]{0.5\linewidth}
		\centering
		\includegraphics[width=\textwidth]{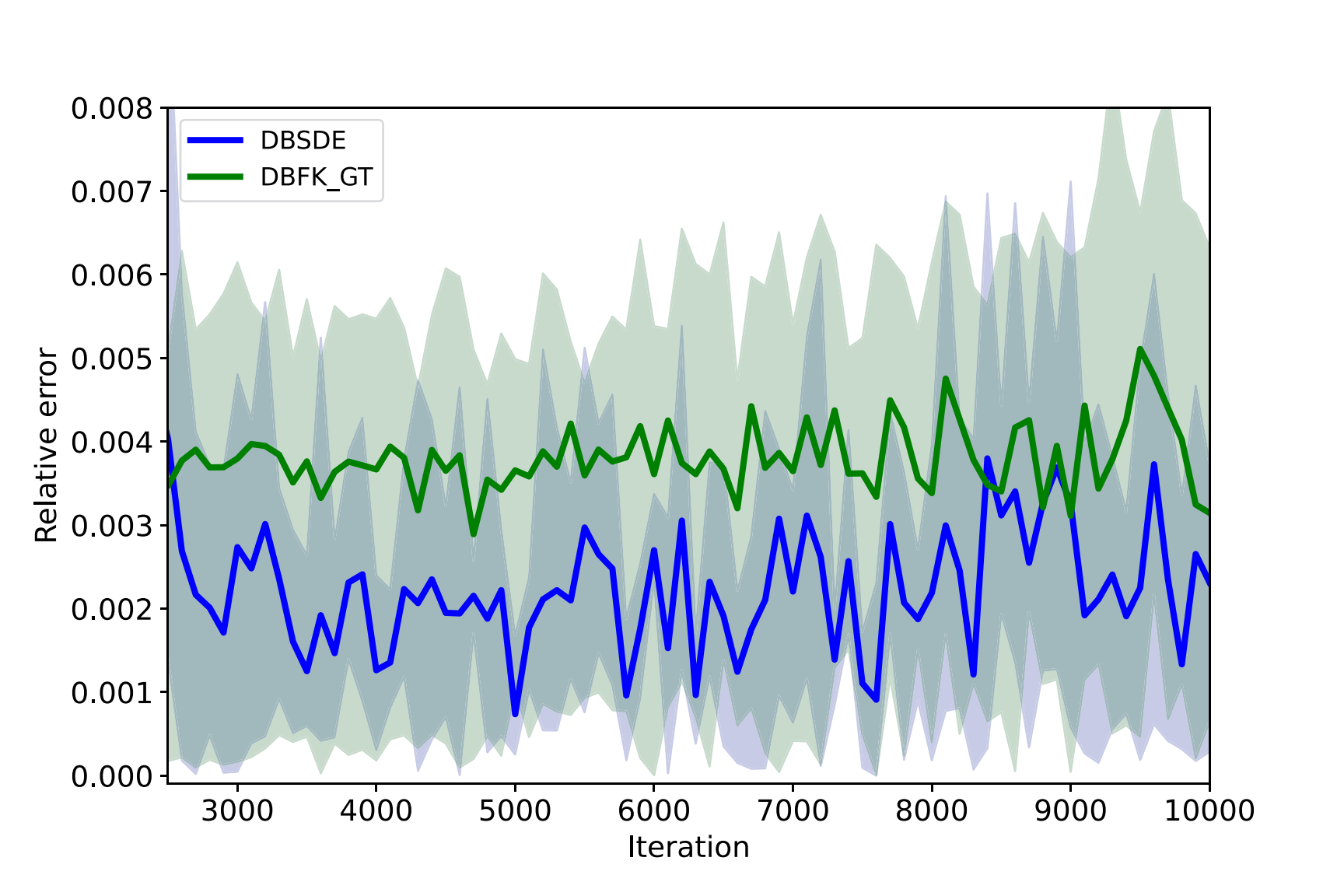}
		\label{fig3:second}
	\end{minipage}
	\vspace{-0.7cm}
	\centering
	\caption{Comparison of Two Methods for Solving the Allen-Cahn Equation}
	\label{figAllencahn}
\end{figure*}

Here, we consider the initial point $\xi=(0,0, \ldots, 0) \in \mathbb{R}^d$, where the spatial dimension is $d=100$ and the terminal time $T=\frac{3}{10}$. Using the branching diffusion method \cite{FENZHI1,FENZHI2,FENZHI3}, we can obtain a reference value for the exact solution of the equation, where $u(0, \xi)=u(0,0, \ldots, 0)\approx 0.052802$.
Similarly, based on the loss function (\ref{TS-FK-LSMC-LOSS-BECK}), (\ref{DS-GT-LOSS}), (\ref{OT-FK-LSMC-LOSS}), and (\ref{BSDE-LOSS}), we set the time step number to $N=20$ and perform independent calculations 5 times. 
The results are reported in Table \ref{biaoallencahn}. 
The DFK-GT method has been observed to substantially improve both the computational efficiency and accuracy, compared to the DS method. This improvement can be attributed to the implementation of a global training scheme and the remodeling of data pairs (\ref{TS-FK-LSMC-LOSS1}).
Moreover, as depicted in Figure \ref{figAllencahn}, in comparison to the widely popular DBSDE method, the DFK-GT method not only ensured higher accuracy but also demonstrated a substantial improvement in computational efficiency.

\begin{table}
	\centering
	\caption{Numerical Results for Allen-Cahn Equation}
	\resizebox{0.8\textwidth}{!}{
		\begin{tabular}{ccccccccc}
			\hline \hline
			&  & \multicolumn{7}{c}{Number of iteration steps} \\
			\cmidrule{3-9}
			&   & 100 & 200 & 600 & 1000 & 2000 & 5000 & 10000 \\
			\midrule
			\multirow{4}{*}{Runtime} & DBSDE         & 10.6  & 14.8 & 31.4  & 48.4  & 90.2   & 214.2  & 429.8 \\
			& DS   & 18.8  & 23.6 & 41.4 & 59.8       & 106.6    & 244.4 & 471.0 \\
			& DS-GT  & 10.2    & 12.4  & 22    & 31.4    & 55   & 125.6   & 243.6 \\
			& DFK-GT    & 8   & 10.2   & 19 & 27.8   & 49.8  & 115.8  & 225.4  \\
			\midrule
			\multirow{4}{*}{Relative error} & DBSDE  & 5.26 & 4.44   & 2.00  & 7.26e-01 & 2.51e-02   & \textbf{7.36e-04} & 2.31e-03 \\
			& DS  & 4.16e-02  & 3.00e-03  & 2.91e-03  & 3.90e-03     & 9.85e-03  & 4.65e-03 & 4.23e-03 \\
			& DS-GT  & 9.99e-01  & 1.02  & 9.83e-01  & 9.74e-01   & 9.53e-01  & 2.41e-01 & 2.50e-03  \\
			& DFK-GT  & 2.15e-01 & 1.45e-01 & 1.08e-02  & 4.23e-03 & 3.80e-03 & 3.65e-03   & 2.93e-03 \\
			\midrule
			\multirow{4}{*}{Loss function} & DBSDE  & 3.55e-02  & 2.32e-02  & 3.46e-03    & 4.77e-04   & 1.93e-04  & 4.56e-05    & 2.64e-05 \\
			& DS  & 3.24e-04  & 4.87e-07   & 1.49e-07 & 4.53e-08   & 6.20e-07  & 4.59e-08  & 7.24e-08 \\
			& DS-GT  & 1.11e+01  & 5.73 & 4.56e-01  & 8.87e-02   & 1.16e-02 & 5.44e-04 & 3.27e-05  \\
			& DFK-GT & 6.44 & 3.51  & 1.76e-01 & 3.74e-03   & 5.82e-04    & 5.88e-04 & 6.43e-04  \\
			\hline \hline
		\end{tabular}
	}
	\label{biaoallencahn}
\end{table}

Importantly, it should be noted that a consistent relationship exists between variations in the loss function and the accuracy of the methods.
Consequently, when the loss function exhibits limited fluctuations (i.e., the loss function ceases to decrease), it becomes possible to ascertain the optimal number of iterations necessary for algorithm convergence.
The results are reported in Table \ref{biaoallencahn2}. It turns out that all the methods provide quite accurate results with the highest computational efficiency for the DFK-GT.

\begin{table*}
	\caption{Iteration Count at Loss Convergence for the Allen-Cahn Equation}
	\centering
	\resizebox{0.6\textwidth}{!}{
		\begin{tabular}{ccccc}
			\hline \hline
			Methods  & Iterations & Relative error & Runtime & Standard deviation  \\
			\hline
			DBSDE  & 6700 &3.98e-03 & 290.2 & 1.05e-04 \\
			\hline
			DS & 800 & 3.04e-03 & 50.6 & 1.65e-04 \\
			\hline
			DS-GT  & 7600 & 5.84e-03 & 187 & \textbf{8.16e-05} \\
			\hline
			DFK-GT   & 1600 & 3.72e-03 & \textbf{41.4} & 2.45e-04 \\
			\hline \hline
		\end{tabular}}
		\label{biaoallencahn1}
\end{table*}

Additionally, by recording the experimental results where the relative error first falls below \(10^{-2}\), we can ascertain the ideal number of iterations required for algorithm convergence. The corresponding results are presented in Table \ref{biaoallencahn2}.
\begin{table*}
	\caption{Iteration Count for the Allen-Cahn Equation (Relative Error Below \(10^{-2}\))}
	\centering
	\resizebox{0.6\textwidth}{!}{
		\begin{tabular}{ccccc}
			\hline \hline
			Methods  & Iterations & Relative error & Runtime & Standard deviation \\
			\hline
			DBSDE  & 1800 &9.45e-03 & 86.8 & 4.42e-03 \\
			\hline
			DS & 200 & 3.00e-03 & 23.6 & 1.29e-03 \\
			\hline
			DS-GT  & 7200 &  9.43e-03 & 177.4 & 1.32e-04 \\
			\hline
			DFK-GT   & 700 & 5.58e-03 & \textbf{20.2} & \textbf{3.13e-04} \\
			\hline \hline
		\end{tabular}
		}
	\label{biaoallencahn2}
\end{table*}



\subsection{Pricing of European financial derivatives with different interest rates for borrowing and lending(PricingDiffrate) equation}
There are various extensions of the traditional linear Black-Scholes equation that incorporate nonlinear phenomena, including transaction costs, default risk, and Knightian uncertainty. In this section, we apply the methods for solving the special nonlinear Black-Scholes equations.
It describes the pricing problem of an European financial derivative in a financial market where the risk-free bank account utilized for hedging purposes exhibits disparate interest rates for borrowing and lending \cite{Bergman4}.

Referring to the general form of a semi-linear parabolic PDE(\ref{bxxpwxpde}),  let $\bar{\mu}=0.06$, $\bar{\sigma}=0.2$, $R^l=0.04$, $ R^b=0.06$, and assume for all $s,t\in[0,T]$, $x=(x_1,\ldots,x_d)\in\mathbb{R}^d$, $y\in\mathbb{R}$, and $z\in\mathbb{R}^{d}$, $d=100$, $T=1/2$, $N=20$, $\mu(t,x)=\bar{\mu}x$,  $\sigma(t, x)=\bar{\sigma}x$ and $\xi=(100,100,\ldots,100)\in\mathbb{R}^d$.
Meanwhile, we choose a terminal condition $g(x)$ and a non-linear term $f(t,x,y,z)$ for the equation:
\begin{equation}\label{EPDFeqg}
	\begin{array}{l}
		g(x) = \max \left\{ {\left[ {{{\max }_{1 \le i \le 100}}{x_i}} \right] - 120,0} \right\} - 2\max \left\{ {\left[ {{{\max }_{1 \le i \le 100}}{x_i}} \right] - 150,0} \right\},
	\end{array}
\end{equation}
\begin{equation}\label{EPDFeqf}
	\begin{array}{l}
		f(t,x,y,z) =  - {R^l}y - \frac{{\left( {\bar \mu  - {R^l}} \right)}}{{\bar \sigma }}\sum\limits_{i = 1}^d {{z_i}}  + \left( {{R^b} - {R^l}} \right)\max \left\{ {0,\left[ {\frac{1}{{\bar \sigma }}\sum\limits_{i = 1}^d {{z_i}} } \right] - y} \right\},
	\end{array}
\end{equation}
At this point, the equation can be represented on the region $t \in [0, T)$ and $x \in \mathbb{R}^d$:
\begin{equation}\label{EPDFeq}
	{\begin{array}{*{20}{l}}
			{\frac{{\partial u}}{{\partial t}}(t,x) + \frac{{{{\bar \sigma }^2}}}{2}\sum\limits_{i = 1}^d {{{\left| {{x_i}} \right|}^2}} \frac{{{\partial ^2}u}}{{\partial x_i^2}}(t,x)}\\
			{ + \max \left\{ {{R^b}\left( {\left[ {\sum\limits_{i = 1}^d {{x_i}} \left( {\frac{{\partial u}}{{\partial {x_i}}}} \right)(t,x)} \right] - u(t,x)} \right),{R^l}\left( {\left[ {\sum\limits_{i = 1}^d {{x_i}} \left( {\frac{{\partial u}}{{\partial {x_i}}}} \right)(t,x)} \right] - u(t,x)} \right)} \right\} = 0.}
	\end{array}}
\end{equation}
The PDE (\ref{EPDFeq}) can also be equivalently expressed as:
\begin{equation}\label{EPDFeq1}
	\begin{array}{l}
		\frac{{\partial u}}{{\partial t}}(t,x) + \frac{{{{\bar \sigma }^2}}}{2}\sum\limits_{i = 1}^d {{{\left| {{x_i}} \right|}^2}} \frac{{{\partial ^2}u}}{{\partial x_i^2}}(t,x)\\
		- \min \left\{ {{R^b}\left( {u(t,x) - \left[ {\sum\limits_{i = 1}^d {{x_i}} \left( {\frac{{\partial u}}{{\partial {x_i}}}} \right)(t,x)} \right]} \right),{R^l}\left( {u(t,x) - \left[ {\sum\limits_{i = 1}^d {{x_i}} \left( {\frac{{\partial u}}{{\partial {x_i}}}} \right)(t,x)} \right]} \right)} \right\} = 0.
	\end{array}
\end{equation}
	
\begin{figure*}
	\centering
	\begin{minipage}[t]{0.5\linewidth}
			\centering
			\includegraphics[width=\textwidth]{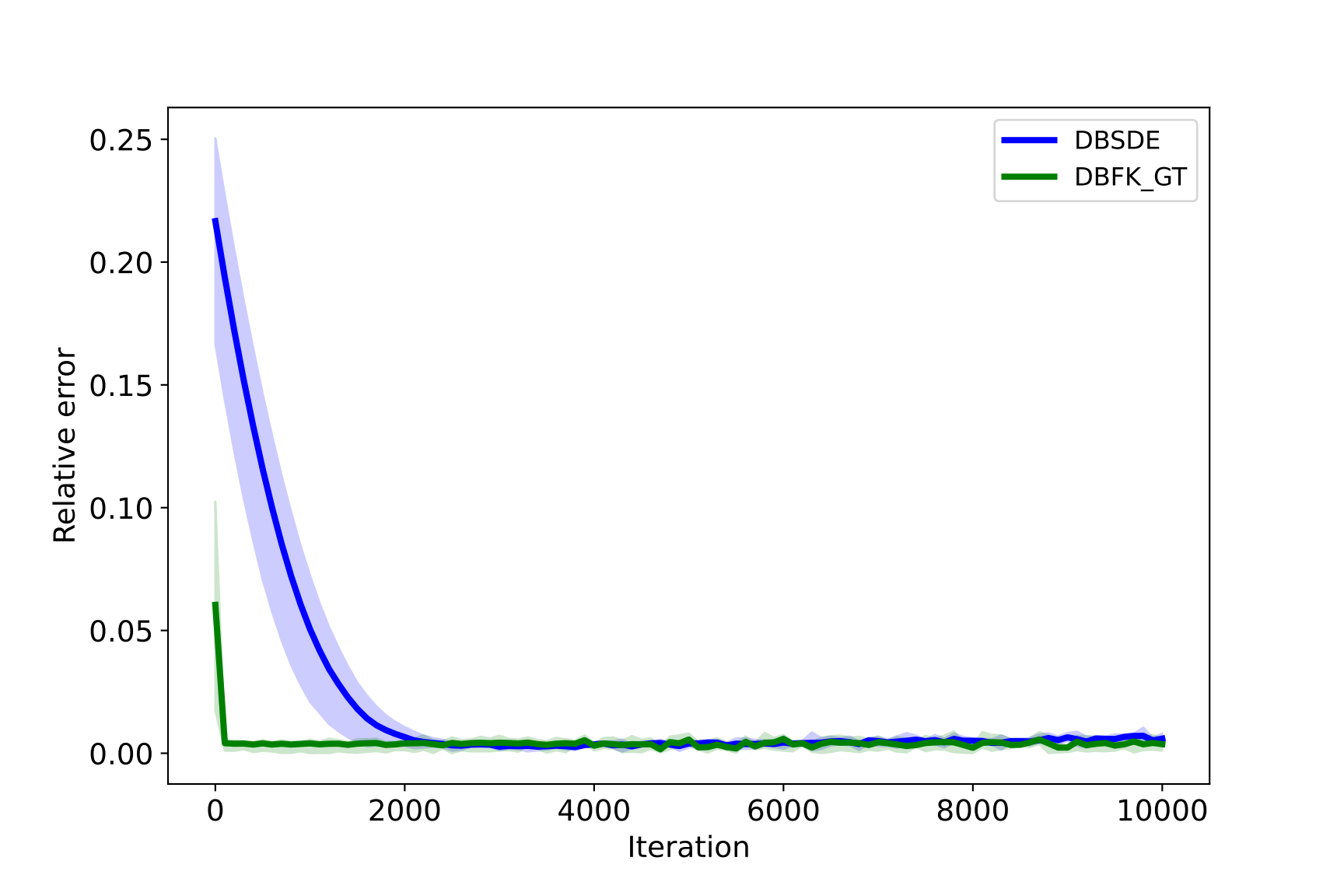}
			\label{fig4:first}
		\end{minipage}%
		\hfill
		\begin{minipage}[t]{0.5\linewidth}
			\centering
			\includegraphics[width=\textwidth]{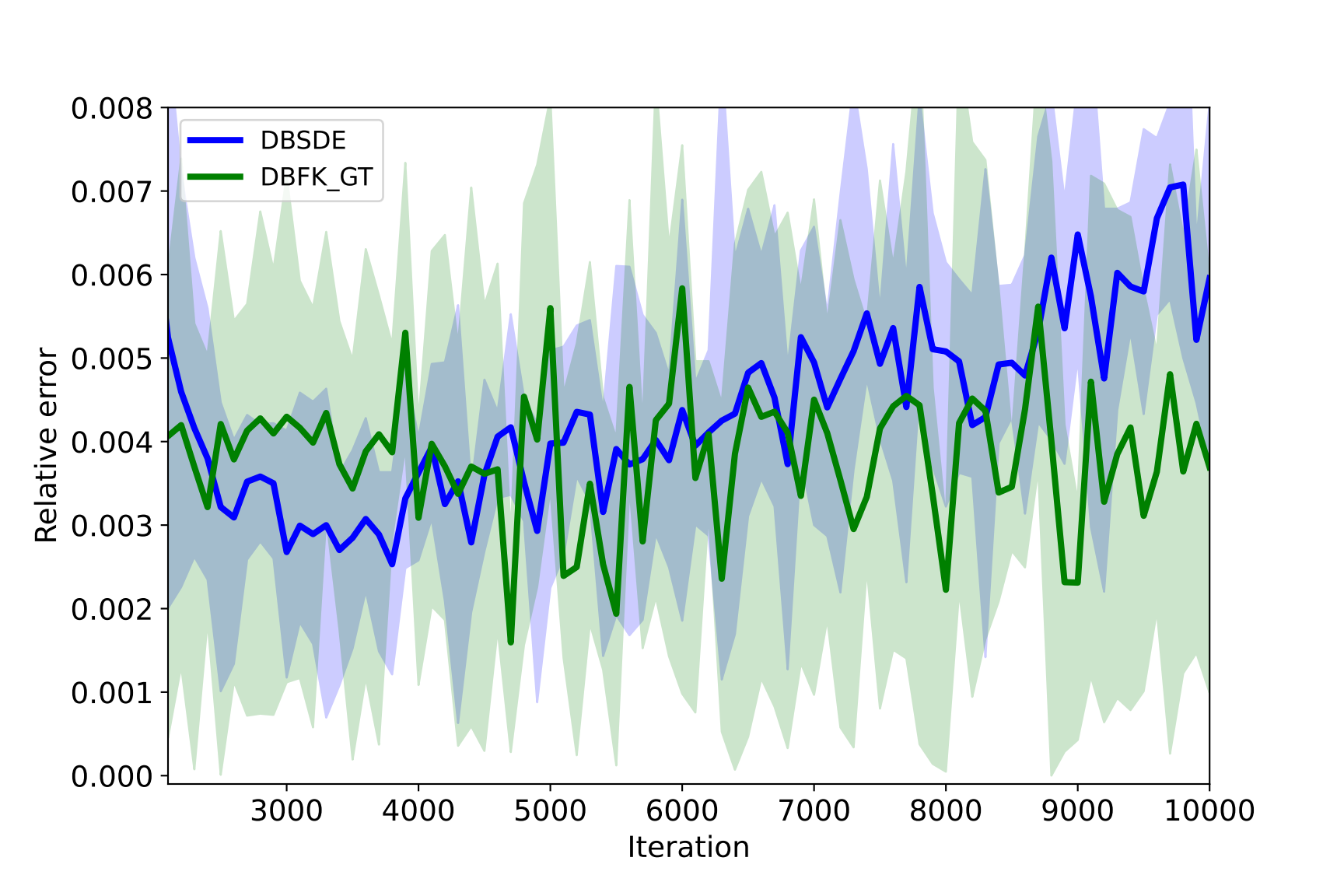}
			\label{fig4:second}
		\end{minipage}
		\vspace{-0.7cm}
		\centering
		\caption{Comparison of Two methods for Solving the PricingDiffrate Equation}
		\label{figdiffrate}
\end{figure*}

The solution of the equation can be obtained through the Multilevel-Picard approximation method \cite{multilevel-Picard}, which gives a value of $21.299$. 
Based on the loss function (\ref{TS-FK-LSMC-LOSS-BECK}), (\ref{DS-GT-LOSS}), (\ref{OT-FK-LSMC-LOSS}), and (\ref{BSDE-LOSS}),  we set the time step number to $N=20$ and perform independent calculations 5 times. 
The results are reported in Table \ref{biaodiffrate}. 
Specifically, we observed a significant advantage of the DFK-GT method over the DS method in terms of computational efficiency and accuracy. This improvement can be attributed to the implementation of a global training scheme and the remodeling of data pairs (\ref{TS-FK-LSMC-LOSS1}).
Additionally, Figure \ref{figdiffrate} and table \ref{biaodiffrate} clearly illustrate the faster convergence speed of the DFK-GT method compared to the DBSDE method, while maintaining a similar level of robustness.

\begin{table}[htbp]
	\centering
	\caption{Numerical Results for PricingDiffrate Equation}
	\resizebox{0.8\textwidth}{!}{
		\begin{tabular}{ccccccccc}
			\hline \hline
			&  & \multicolumn{7}{c}{Number of iteration steps} \\
			\cmidrule{3-9}
			&   & 100 & 200 & 600 & 1000 & 2000 & 5000 & 10000 \\
			\midrule
			\multirow{4}{*}{Runtime} & DBSDE & 11.6 & 16 & 33.6 & 51.6 & 95.8 & 228.6 & 451.6 \\
			& DS  & 20.0 & 23.0 & 35.2 & 48.4 & 81.2 & 188.4 & 365.8 \\ 
			& DS-GT & 10.8 & 13.2 & 24 & 34.4 & 60.6 & 138.4 & 268.8  \\
			& DFK-GT & 9 & 11.2 & 21.4 & 31.4 & 56.2 & 130.4 & 254.4  \\
			\midrule
			\multirow{4}{*}{Relative error} & DBSDE  & 1.94e-01 & 1.72e-01 & 9.99e-02 & 5.05e-02 & 6.57e-03 & 3.98e-03 &  5.96e-03 \\
			& DS  & 7.42e-01 & 5.79e-01 & 3.89e-02 & 4.26e-03 & 3.18e-02 & 2.07e-02 & 2.21e-02 \\
			& DS-GT  & 1.17 & 4.94e-01 & 3.41e-02 & 2.85e-02 & 4.28e-02 & 2.84e-02 & 1.08e-01 \\
			& DFK-GT  & 4.09e-03 & 3.88e-03 & 3.54e-03 & 4.01e-03 & 4.01e-03 & 5.60e-03 & \textbf{2.69e-03} \\
			\midrule
			\multirow{4}{*}{Loss function} & DBSDE  & 5.12e+01 & 4.74e+01 & 3.79e+01 & 3.41e+01 & 3.26e+01 & 3.26e+01 & 3.25e+01 \\
			& DS  & 5.53e-02 & 5.21e-02 & 6.46e-02 & 5.08e-02 & 5.08e-02 & 4.45e-02 & 5.59e-02 \\ 
			& DS-GT  & 7.06e+03 & 6.27e+03 & 3.39e+03  & 1.70e+03 & 4.74e+02 & 2.53e+02 & 4.70e+02 \\
			& DFK-GT  & 4.28e+03 & 3.91e+03 & 2.39e+03 & 1.50e+03 & 8.19e+02 & 7.93e+02 & 6.27e+02 \\
			\hline \hline
		\end{tabular}
	}
	\label{biaodiffrate}
\end{table}


Importantly, a consistent relationship is observed between variations in the loss function and the accuracy of the methods. Specifically, when the loss function exhibits limited fluctuations (i.e., the loss function ceases to decrease), the optimal number of iterations required for algorithm convergence can be determined. The corresponding results are summarized in Table \ref{biaodiffrate1}. Notably, all methods yield highly accurate results, with DFK-GT achieving the highest computational efficiency and the lowest standard deviation.

\begin{table}
	\caption{Iteration Count at Loss Convergence for the PricingDiffrate Equation}
	\centering
	\resizebox{0.6\textwidth}{!}{
		\begin{tabular}{ccccc}
			\hline \hline
			Methods  & Iterations & Relative error &  Runtime & Standard deviation \\
			\hline
			DBSDE  & 2200 & 4.60e-03 & 106.4 & 4.77e-02\\
			\hline
			DS & 2000 & 3.18e-02 & 81.2 &  8.90e-01 \\
			\hline
			DS-GT  & 2800 & 5.58e-02 & 81 & 1.37 \\
			\hline
			DFK-GT   & 1800 & 2.45e-03 & \textbf{51.6} & \textbf{2.17e-02} \\
			\hline \hline
		\end{tabular}}
		\label{biaodiffrate1}
\end{table}

Additionally, by recording the experimental results where the relative error first reaches the $10^{-2}$ level, we can ascertain the ideal number of iterations required for algorithm convergence. The corresponding results are presented in Table \ref{biaodiffrate2}.
\begin{table}
	\caption{Iteration Count for the PricingDiffrate Equation (Relative Error Below \(10^{-2}\))}
	\centering
	\resizebox{0.6\textwidth}{!}{
		\begin{tabular}{ccccc}
			\hline \hline
			Methods  & Optimal Iterations & Relative error & Runtime & Standard deviation \\
			\hline
			DBSDE  & 2300 &6.79e-03 & 102.4 & 3.21e-02 \\
			\hline
			DS & 1000 & 4.26e-03 & 48.4 & 3.77e-01 \\
			\hline
			DS-GT  & 7700 &  9.86e-03 & 208.8 & 2.25e-01 \\
			\hline
			DFK-GT   & 100 & 4.09e-03 & \textbf{9} & 7.64e-02 \\
			\hline \hline
		\end{tabular}
		}
	\label{biaodiffrate2}
\end{table}

\section{Conclusions}
In this paper, we introduced a novel training approach for solving high-dimensional semilinear parabolic partial differential equations (PDEs) using neural networks. Our method incorporates two key features: (1) a global training strategy that updates the neural network across all time steps simultaneously, rather than in a sequential step-by-step manner, and (2) a new data pairs generation approach that ensures consistency with the direct Monte Carlo scheme when applied to linear parabolic PDEs.

Through numerical experiments, we compare the Deep Splitting (DS) method and the Deep Splitting method under Global Training (DS-GT). The results indicate that global training improves computational speed compared to the step-by-step training approach. Furthermore, by comparing the Deep Splitting method under Global Training (DS-GT) with the Deep Feynman-Kac method under Global Training (DFK-GT), we observe that remodeling the data pairs enhances both computational accuracy and efficiency.

\begin{figure*}[!hbtp]
	\begin{center}
		\includegraphics[angle=0,width=6in]{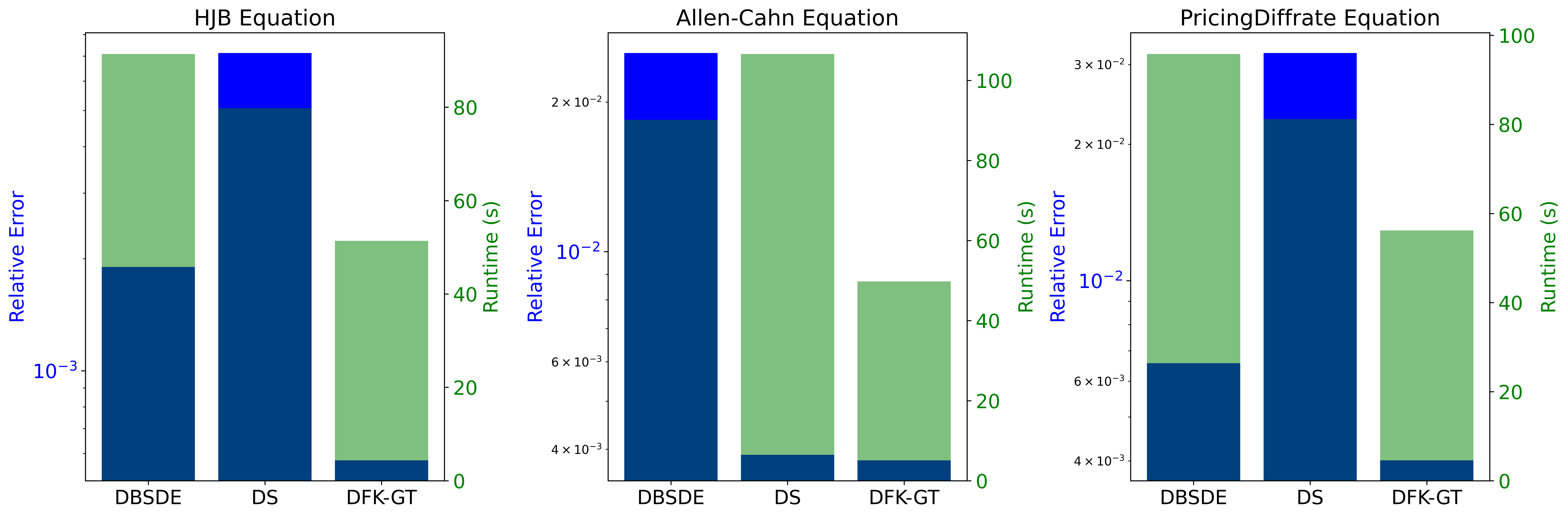}		
	\end{center}
	\caption{Comparison of Runtime and Accuracy on Different Numerical Examples (2000 Iterations)}
	\label{fig_compare}
\end{figure*}

Overall, our findings demonstrate that the proposed approach (DFK-GT) significantly improves both computational efficiency and solution accuracy compared to the Deep Splitting (DS) method and the Deep BSDE (DBSDE) method. The results for a training limit of 2000 iterations are shown in Figure \ref{fig_compare}.

\clearpage
\bibliography{main}
\bibliographystyle{siam}

\appendix
\section{Appendix: Deep BSDE Method} \label{Appendix:DBSDE}
In recent years, there has been a surge of interest in using neural networks to solve high-dimensional PDEs and BSDEs. In particular, Weinan E et al. proposed the Deep BSDE method \cite{E2017,E2018}.

After adapting the solution of PDE (\ref{bxxpwxpde}) to the SDE (\ref{BSDEeq}) by BSDE theory, the next step is to discretize the SDE in  time. Given a partitioning of the interval $[0,T]:0=t_0<t_1<\dots<t_N=T$ with $\Delta t_n=t_{n+1}-t_n$, the solution at each time step can be approximated using the Euler-Maruyama scheme.
\begin{equation}\label{ls-BSDEeq}
	\begin{aligned}
		u\left( {{t_{n + 1}},{X_{{t_{n + 1}}}}} \right) - u\left( {{t_n},{X_{{t_n}}}} \right)  & \approx  - f\left( {{t_n},{X_{{t_n}}},u\left( {{t_n},{X_{{t_n}}}} \right),{\sigma ^{{T}}}\left( {{t_n},{X_{{t_n}}}} \right)\nabla u\left( {{t_n},{X_{{t_n}}}} \right)} \right)\Delta {t_n}\\
		& + {{\left[ {\nabla u\left( {{t_n},{X_{{t_n}}}} \right)} \right]}^{\rm{T}}}\sigma \left( {{t_n},{X_{{t_n}}}} \right)\Delta {W_n},n = 0,1,...,N - 1.
	\end{aligned}
\end{equation}
To achieve a globally approximate scheme, neural networks can be incorporated into the forward discretization process. The first step towards this is to obtain training data by sampling $M$ independent paths ${\left\{ {X_{{t_n}}^m} \right\}_{0 \le n \le N,m = 1,2, \ldots, M}}$, where $\left\{X_{t_0}^m\right\}_{m=1, \ldots, M}=\xi$.The critical step next is employing the neural network parameters ${\theta _{{u_0}}}$, ${\theta _{\nabla {u_0}}}$ to approximate the solution and the gradient function at $t=t_0$ respectively. Let $\theta_n$ represent all network parameters approximating the gradient function $\nabla u(t,x)$ by the neural network at time $t=t_n$, where $n = 1,2,..., N - 1$.  
With the above approximations, the total set of parameters is $\theta=\left\{{\theta _{{u_0}}}, {\theta _{\nabla {u_0}}}, \theta_1, \theta_2, \ldots, \theta_{N-1}\right\}$. The equation (\ref{ls-BSDEeq}) can be rewritten as follows.
\begin{equation}\label{nn-ls-BSDEeq}
	\begin{aligned}
		\hat u\left( {{t_{n + 1}},X_{{t_{n + 1}}}^m} \right) & - \hat u\left( {{t_n},X_{{t_n}}^m} \right)  \approx  - f\left( {{t_n},X_{{t_n}}^m,\hat u\left( {{t_n},X_{{t_n}}^m} \right),{\sigma ^{{T}}}\left( {{t_n},X_{{t_n}}^m} \right)\nabla u_{{t_n}}^\theta \left( {{t_n},X_{{t_n}}^m} \right)} \right)\Delta {t_n}\\
		& + {\left[ {\nabla u_{{t_n}}^\theta \left( {{t_n},X_{{t_n}}^m} \right)} \right]^{\rm{T}}}\sigma \left( {{t_n},X_{{t_n}}^m} \right)\Delta W_{{t_n}}^m,m = 1,2,...,M,n = 0,1,...,N - 1.
	\end{aligned}
\end{equation}
In particular, when $t=t_0$, we have
\begin{equation}\label{t0-nn-ls-BSDEeq}
	\begin{aligned}
		\hat u\left( {{t_1},X_{{t_1}}^m} \right) \approx {\theta _{{u_0}}} - f\left( {{t_0},\xi ,{\theta _{{u_0}}},{\sigma ^T}\left( {{t_0},\xi } \right){\theta _{\nabla {u_0}}}} \right)\Delta {t_0} + {\left[ {{\theta _{\nabla {u_0}}}} \right]^{\rm{T}}}\sigma \left( {{t_0},\xi } \right)\Delta W_{{t_0}}^m,m = 1,2,...,M.
	\end{aligned}
\end{equation}

By applying the globally approximate scheme (\ref{nn-ls-BSDEeq}), an approximate value of $u\left(t_N, X_{t_N}^m\right)$ denoted as $\hat{u}\left(t_N, X_{t_N}^m\right)$ can be obtained, where $m=1,2, \ldots, M$. The matching of a given terminal condition can define the expected loss function.
\begin{equation}\label{BSDE-LOSS}
	\begin{aligned}
		l(\theta ) = \frac{1}{M}\sum\limits_{m = 1}^M {{{\left| {g\left( {{X_T}} \right) - \hat u\left( {{t_N},X_{{t_N}}^m} \right)} \right|}^2}} .
	\end{aligned}
\end{equation}
Network parameters can be trained by the optimizer, such as SGD \cite{SGD}, L-BFGS-B \cite{LBFGSB1}\cite{LBFGSB2}, Adam \cite{ADAM}, and Adagrad \cite{ADAGRAD} algorithms. Through this process, it becomes evident that an approximate solution
${\theta _{{u_0}}}$ for $u(0, \xi)$ can be obtained. In the work of Weinan E et al., they opted to use the Adam algorithm.


\begin{Remark}
According to the Feynman-Kac formula, the solution to the PDE (\ref{bxxpwxpde}), denoted as $u(t,x)$, can be expressed as the conditional expectation (\ref{F-Keq}). By partitioning the time interval, we have
\begin{equation}\label{F-Keq1}
			\begin{aligned}
				u({t_n},{x}) = \mathbb{E}\left[ {u({t_{n + 1}},{X_{{t_{n + 1}}}}) + \int_{{t_n}}^{{t_{n + 1}}} \!\!f \left( {s,{X_s},u\left( {s,{X_s}} \right),{\sigma ^T}(s,{X_s})\nabla u\left( {s,{X_s}} \right)} \right){\rm{d}}s{\rm{\mid }}{X_{{t_n}}} = x } \right].
		\end{aligned}
\end{equation}
Here, the $d$-dimensional stochastic process $\left\{ X_s \right\}_{s \in [{t_n},{t_{n+1}}]}$ satisfies the SDE (\ref{2.7}). 
When performing quadrature on integrals, we can estimate the values of neighboring functions for $n = 0,1,..., N - 1$. 
\begin{equation}\label{deep-KF-eq}
    \begin{aligned}
			u\left( {{t_{n+1}},{X_{{t_{n+1 }}}}} \right)  \approx u\left( {{t_{n }},{X_{{t_{n}}}}} \right) - f\left( {{t_n},{X_{{t_n}}},u\left( {{t_n},{X_{{t_n}}}} \right),{\sigma ^{{T}}}\left( {{t_n},{X_{{t_n}}}} \right)\nabla u\left( {{t_n},{X_{{t_n}}}} \right)} \right)\Delta {t_n}.
		\end{aligned}
\end{equation}
By treating equation (\ref{deep-KF-eq}) in a similar way to equation (\ref{ls-BSDEeq}) from the Deep BSDE method, we can develop a new approach so-called as the Deep forward Feynman-Kac method. Both methods share the same form of network parameters and loss function, but they differ in the treatment of the Brownian motion term ${\left[ {\nabla u\left( {{t_n},{X_{{t_n}}}} \right)} \right]^{\rm{T}}}$ 
$\sigma \left( {{t_n},{X_{{t_n}}}} \right)\Delta {W_n}$.
\end{Remark}

\begin{figure*}[!hbtp]
	\begin{center}
		\includegraphics[angle=0,width=6in]{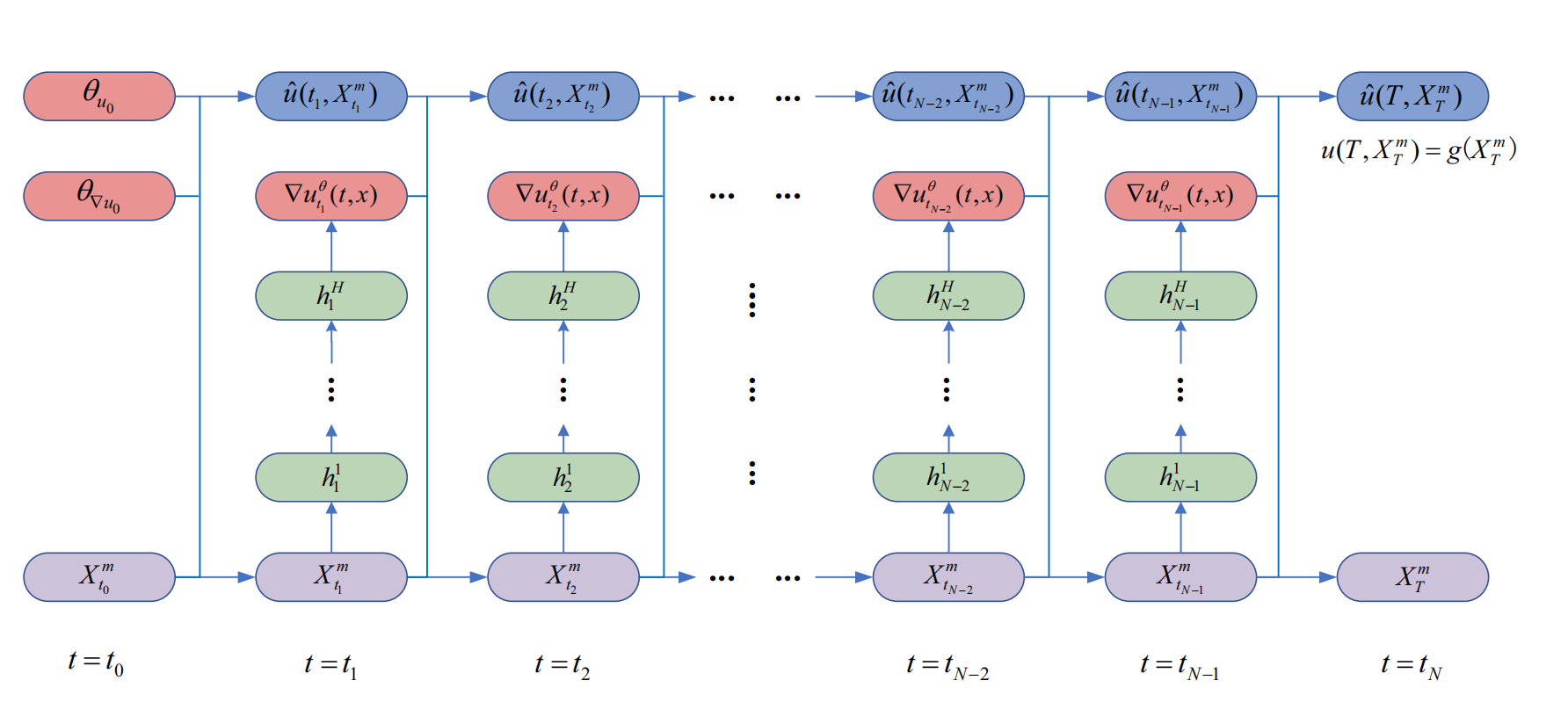}		
	\end{center}
	\vspace{-0.7cm}
	\caption{The neural network architecture for Deep BSDE method. The network consists of multiple sub-networks, with each sub-network corresponding to a time interval. Each sub-network has $H$ hidden layers, where $H$ is a user-defined hyperparameter. At each time layer $t = t_n$, the intermediate neurons of the sub-network are represented as $h_1^n$, $h_2^n$, ..., $h_H^n$, where $n$ ranges from $1$ to $N-1$. It should be noted that in addition to these, ${\theta _{{u_0}}}$ and ${\theta _{\nabla {u_0}}}$ are also network parameters that need to be optimized.}
	\label{fig0}
\end{figure*}

\end{document}